\documentclass[preprint,aps]{revtex4}
\usepackage{amsmath}
\usepackage{graphicx}

\def\be{\begin{equation}}
\def\ee{\end{equation}}
\def \bea#1\eea {\begin{eqnarray}#1\end{eqnarray}}

\begin{document}

\title{
Structural Order for One-Scale and Two-Scale Potentials
}

\author{Zhenyu Yan$^1$, Sergey V. Buldyrev$^{2,1}$, Nicolas
Giovambattista$^{3}$, and H. Eugene Stanley$^1$}

\affiliation{
$^1$Center for Polymer Studies and Department of Physics, Boston 
University, Boston MA 02215, USA\\
$^2$Department of Physics, Yeshiva University, 500 West 185th Street,
  New York, NY 10033 USA\\
$^3$Department of Chemical Engineering, Princeton University, 
Princeton, New Jersey 08544-5263 USA
}

\date{LE10704 ~~~ybgs.tex ~~~ Received 27 May 2005 ~~~ Revised \today }

\begin{abstract}

We perform molecular dynamics simulations to investigate the
relationship between structural order and water-like dynamic and
thermodynamic anomalies in spherically-symmetric
potentials having either one or two characteristic length scales.
Structural order is characterized by translational and orientational
order parameters.
We find that (i) dynamic and thermodynamic anomalies exist for both one-scale
and two-scale ramp potentials, and (ii) water-like structural order
anomalies exist only for the two-scale ramp potential.  Our findings
suggest that the water-like relationship between structural order
and anomalies is related to the presence of two different length
scales in the potential.

\end{abstract}


\maketitle

Most liquids become denser upon cooling and more viscous upon
compression. However, water and many other
liquids with local tetrahedral order \cite{angellPCCP} (e.g., silica, silicon,
carbon, and phosphorous) show a decrease in density upon cooling
(density anomaly) and an increase of diffusivity upon pressurizing
(diffusion anomaly).  These liquids share many other thermodynamical
anomalous properties than those mentioned above. For instance,
experiments in phosphorous indicate the presence of a liquid-liquid
phase transition \cite{phos} and similar results are obtained from
computer simulations in silica \cite{saika}, silicon \cite{sri}, and
water \cite{poole}.  A possible explanation of these anomalies
is the tendency of these liquids to form bonds resulting in local open
structures not present in simple liquids.  Therefore, much effort has
been expended to understand the relationship between the structure and
the dynamical and thermodynamic anomalies observed in tetrahedral
liquids.

Simple spherically-symmetric potentials with a `core-softened'
or a repulsive interaction at short distances are able to generate
water-like density and diffusion anomalies
\cite{pablo91,stillinger97,Jagla99,Sadr98,Scala01,Stell72,franzese,predpreprint},
and can even show a liquid-liquid transition
\cite{franzese}. These findings in simple liquids imply that strong
orientational interactions (as observed in water or silica) are not a
{\it necessary\/} condition for a liquid to have thermodynamic and
dynamical anomalies.  It is still not clear whether such strong
orientational interactions are necessary for a liquid to have structural
anomalies. Here we address the question of whether spherically-symmetric
potentials are also able to reproduce the structural anomalies found in
systems with local tetrahedral order.

Errington and Debenedetti \cite{jeffrey01} (ED) have studied
the microscopic structural order in liquid water by using simple
geometrical metrics or order parameters.  The structural order was
characterized using two different metrics: a translational order
parameter $t$ \cite{errington03}, quantifying the tendency of particle
pairs to adopt preferential separations, and a bond-orientational order
parameter $q$ \cite{chau98,jeffrey01} quantifying the extent to which a
molecule and its four nearest neighbors arrange in a tetrahedral local
structure, as is the case in hexagonal ice. 
A useful way of 
investigating structural order in liquids is to map state points onto 
the $t-q$ plane. Such a representation was introduced by Torquato and 
coworkers~\cite{torquato00}, who applied it to sphere packings and referred to it 
as an {\it order map}. ED used the order map to investigate structural order 
in water~\cite{jeffrey01}. Because of the distinctive features discovered in that 
study, in what follows we refer to water-like order maps as the {\it ED 
order map}. ED found that the state points accessible to the 
liquid state in the order map
fall into a two-dimensional area, meaning that in general $t$
and $q$ are independent. However, for those state points where the
dynamical and thermodynamical anomalies occur, the $t-q$ parameters fall
on a line in the ED order map, meaning that they are strictly
correlated. This is a clear evidence of the relationship between
structure and water anomalies.  Shell {\it et al.} \cite{shell02} tested
the ED order map in silica
and found it qualitatively similar to water. However, 
$t-q$ points corresponding to anomalies do not fall on a single line as in
water but fall on a stripe region. 

The ED order map was studied for
hard spheres~\cite{torquato00,truskett00} and
Lennard-Jones~\cite{errington03} system. In these systems, where no
dynamical or thermodynamic anomalies are observed, the liquid state
points {\it always} fall on a single line on the ED order map. In other
words, in contrast to the case of silica or water, the $q-t$ parameters
are {\it always} strongly correlated.  
In light of these findings, it is
natural to inquire about the ED order map of systems that are
spherically symmetric and, in addition, exhibit water-like anomalies in
their thermodynamic and transport properties.

\begin{figure}
\includegraphics[width=10cm]{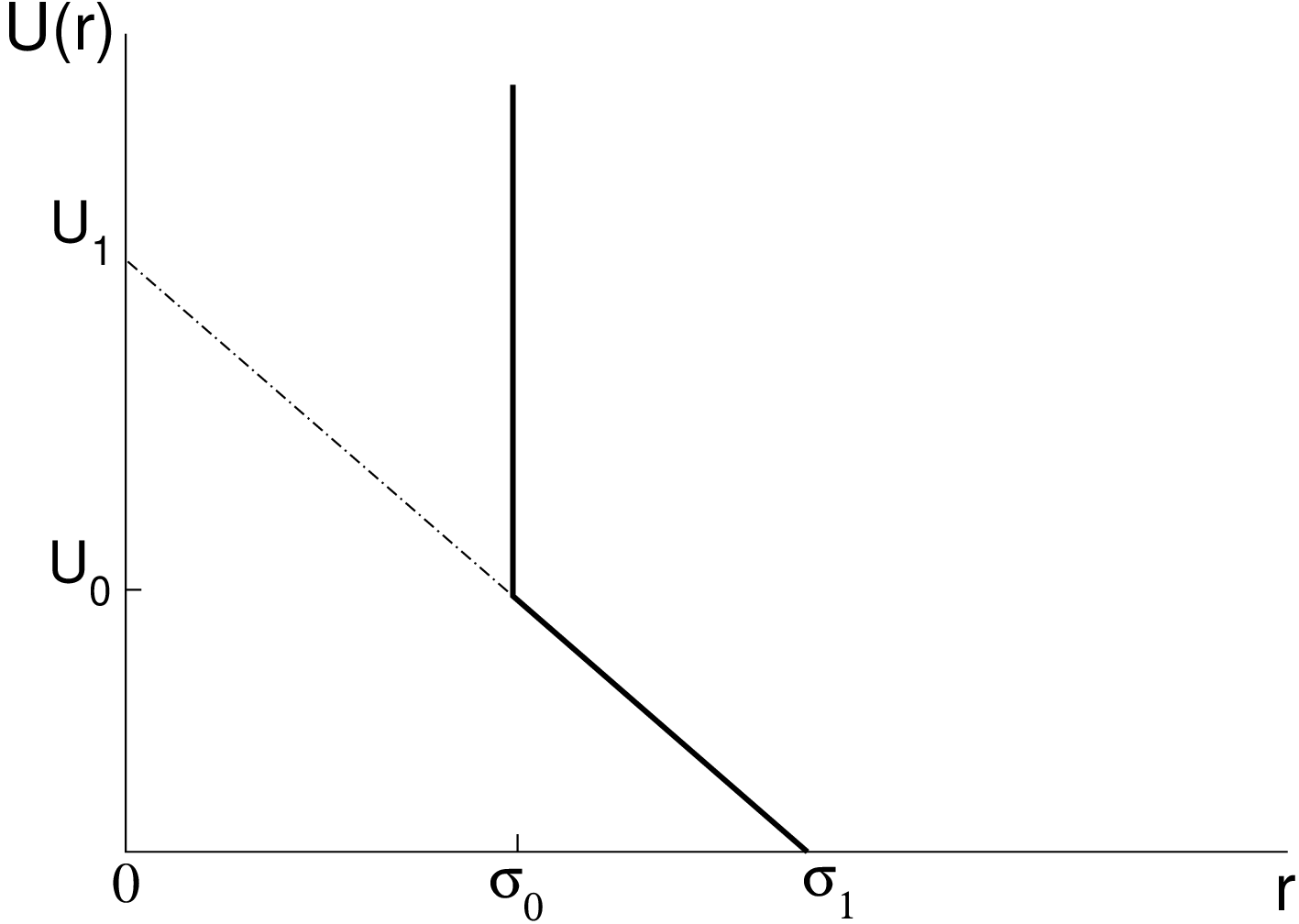}
\caption{The ramp potential introduced by Jagla~\cite{Jagla99}.  
$\sigma_0$ corresponds to the hard-core distance,
$\sigma_1$ characterizes a softer repulsion
range that can be overcome at high pressure.
The 1SRP has $\sigma_0=0$, while the 2SRP
has two length scales with $\sigma_1/\sigma_0=1.76$.}
\label{ramppoten}
\end{figure}

Here we perform discrete molecular dynamics simulations 
to study the
model introduced by Jagla~\cite{Jagla99} (see Fig.~\ref{ramppoten}).
%
%
We will identify the model
with $\lambda\equiv\sigma_1/\sigma_0=1.76$
as the two-scale ramp potential (2SRP) model, and the model with
$\sigma_0=0$ as the one-scale ramp potential
(1SRP) model.
We use NVT ensemble for a system composed by
$N=1728$ (2SRP) or $850$ (1SRP) particles with periodic
boundary conditions and control the temperature with the Berendsen
thermostat.  The details of the simulation are given in
Ref.~\cite{predpreprint}.  However, we note that we use different units
than in Ref.~\cite{predpreprint}: lengths are measured in units of
$\sigma_1$ and energy is measured in units of $U_1$.
We also check that the results do not depend on the number of particles and
the value of $\sigma_1$ and $U_1$ after renormalization.  

We use the translational order parameter~\cite{jeffrey01,shell02,errington03}, 
%
\begin{equation}
t\equiv\int_{0}^{s_c}{\lvert}g(s)-1{\rvert}ds.
\label{top}
\end{equation}
Here $s{\equiv}r\rho_n^{1/3}$ is the radial distance scaled by the
mean interparticle distance, $g(s)$ is the pair correlation function, and $s_c$ a
numerical cutoff that can be set to a suitable value (we choose $s_c$ so
that it corresponds to one-half the simulation box size). For a
completely uncorrelated system, $g(s)\equiv 1$, and thus $t=0$. 
For systems with long-range order, the modulations in $g(s)$ persist over 
large distances, causing $t$ to grow.

An orientational order parameter introduced by
Steinhardt {\it et al.}~\cite{steinhardt83} and used in 
Refs.~\cite{torquato00,truskett00,errington03,huerta04}
is modified to characterize the average local
order of the system~\cite{footnoteDiffinQ6}. 
For each particle, we define
$12$ vectors (or bonds) connecting the central particle with each of its
$12$ nearest neighbors.  Each bond is characterized by two angles
$(\theta,\varphi)$ and the corresponding spherical harmonic $Y_{\ell
m}(\theta,\varphi)$ can be computed. The orientational order parameter
associated with each particle $i$ is
\begin{equation}
Q_{{\ell}i}\equiv\left[\frac{4\pi}{2\ell+1}\sum_{m=-\ell}^{m=\ell}\vert\overline{Y}_{\ell
m}\vert^2 \right]^\frac{1}{2}.
\label{q6}
\end{equation}
Here, $\overline{Y}_{\ell m}(\theta,\varphi)$
denotes the average of $Y_{\ell m}(\theta,\varphi)$
over the 12 bonds associated with particle $i$.
For $\ell=6$~\cite{errington03}, $Q_{\ell}$ has maximum
value for most crystals such as fcc, hcp and bcc~\cite{steinhardt83}.
The values of $Q_{6i}$
for each molecule in the system obey a Gaussian distribution, and the
averaged value of $Q_{6i}$ over all the particles
\cite{footnoteDiffinQ6}, $Q_{6s}$, is used to characterize the local
order of the system.
In general, $Q_{6s}$ grows in value as the crystallinity of a system
increases. For example, in fcc lattice, $Q_{6s}^{\rm fcc}=0.574$, and
for uncorrelated system, $Q_{6s}=1/\sqrt{12}\approx0.289$.  Thus,
$Q_{6s}$ provides a measure of orientational order in the
system. We note that fcc is the structure of the stable crystal at low pressure
in the 1SRP and 2SRP models \cite{predpreprint}.

\begin{figure}
\includegraphics[width=12cm,height=11cm]{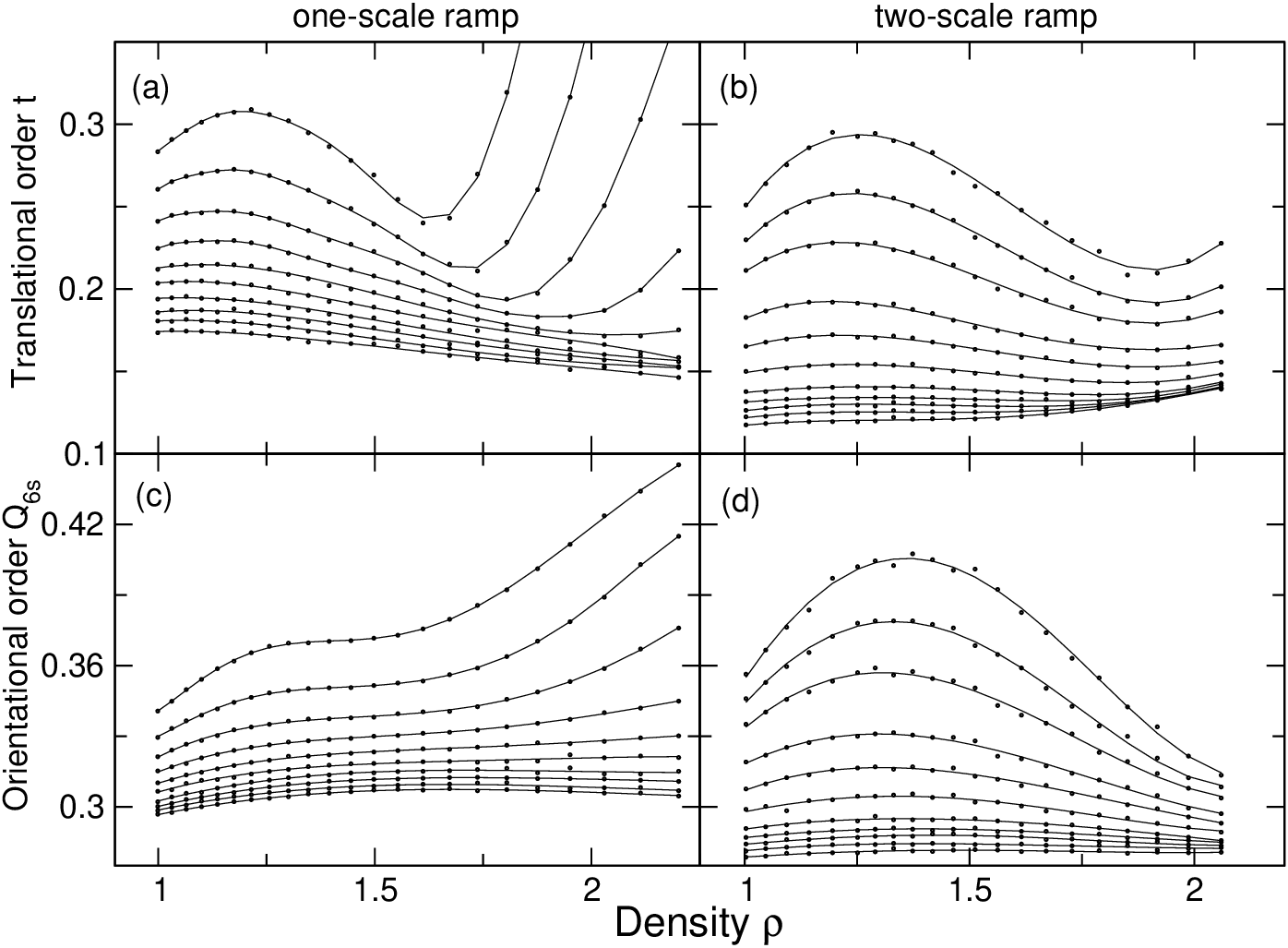}
\caption{The upper panels show the density-dependence of the
     translational order parameter $t$.  The solid lines are polynomial
     fits to the data, introduced as a guide to the eye. (a) One-scale
     ramp potential ($\sigma_0=0$).  From top to bottom isotherms
     correspond to $T=$0.04, 0.05, 0.06, 0.07, 0.08, 0.09, 0.10, 0.11, 0.12, and
     0.13.  (b) Two-scale ramp potential with
     $\sigma_1/\sigma_0=1.76$. From top to bottom isotherms
     correspond to $T=$0.027, 0.036,
     0.045, 0.063, 0.082, 0.109, 0.145, 0.172, 0.2, 0.236, and
     0.290. The lower panels (c)--(d) show the density-dependence of the
     orientational order parameter $Q_{6s}$ for same sets of isotherms.}
\label{qt-d}
\end{figure}

Figs. \ref{qt-d}(a) and \ref{qt-d}(b) show the density dependence of
$t$ for both the 1SRP and 2SRP at different temperatures $T$. The
behavior of $t(\rho)$ is qualitatively the same in both models. At
low-$T$, $t(\rho)$ shows a clear minimum and maximum and, hence, a range
of $\rho$ where $t$ decreases with increasing $\rho$.  This anomaly
implies that the system becomes less ordered upon compression. 
As $T$ increases, the extrema in
$t(\rho)$ disappear. At high-$T$, for the 2SRP, $t$ is a monotonically
increasing function of $\rho$, as is the case in normal
liquids. However, for the 1SRP, $t$ decreases with $\rho$. This behavior
at high $T$ is probably a consequence of the absence of a hard core in
the pair interaction potential. Therefore, upon compression particles
can pass through each other and, as $\rho$ increases, the structure of
the liquid resembles more the structure of the gas.  As expected, 
$t(T)$ and $Q_{6s}(T)$ in both models decreases with $T$ at fixed $\rho$, meaning that
the system becomes more disordered upon heating (Fig.~\ref{qt-d}).

$Q_{6s}$ as a function of density for both models is
shown in Figs.~\ref{qt-d}(c) and \ref{qt-d}(d). At low $T$,
$Q_{6s}(\rho)$ for the 1SRP is a monotonically increasing function of
$\rho$, i.e., orientational order increases upon compression. Instead,
$Q_{6s}(\rho)$ for the 2SRP at low $T$ shows a clear maximum, indicating
a range of density for which $Q_{6s}(\rho)$ decreases with $\rho$. 
We also note that at high pressure/density, the two models have different
crystal structures, hcp for 1SRP and rhombohedral for 2SRP.  
Thus the different $Q_{6s}(\rho)$ in liquid state may be related to the
difference in crystal formation.

\begin{figure}
\includegraphics[width=10cm]{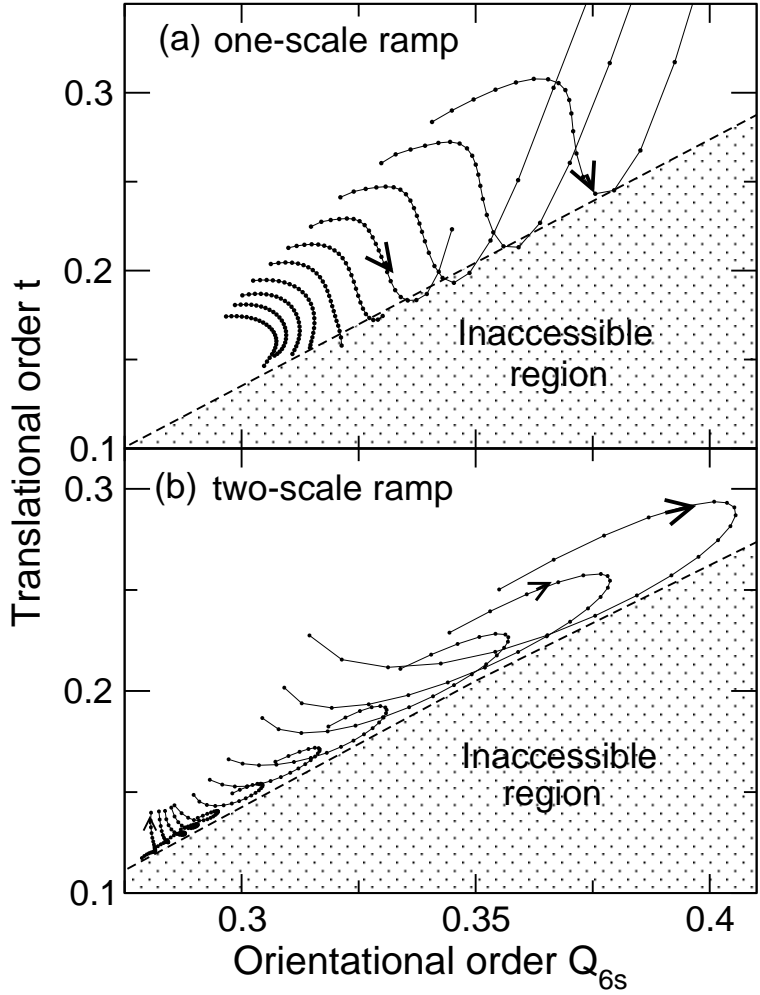}
\caption{(a) The ED order map for the one-scale ramp liquid. Isotherms correspond to $T=$0.04,
  0.05, 0.06, 0.07, 0.08, 0.09, 0.10, 0.11, 0.12, and 0.13 (from top to
  bottom), and arrow indicates the direction of increasing density. 
  For $T\leq 0.08$, $t$ has a maximum at a low density and
  a minimum at a high density but $Q_{6s}$ shows no maximum. 
  (b) The ED order map for the two-scale ramp
  liquid. Isotherms correspond to $T=$0.027, 0.036, 0.045, 0.063, 0.082, 0.109, 0.145, 0.172, 
  0.2, 0.236, and 0.290 (from top to bottom). $Q_{6s}$ and $t$
  have maxima at low densities, and $t$ has minima at high
  densities. We can identify a structurally anomalous
  region bounded by loci of maximum orientational order (at low
  densities) and minimum translational order (at high densities) in
  which both $t$ and $Q_{6s}$ decrease upon compression. 
  All state points within the structurally anomalous region fall in a narrow stripe
  region adjacent to the inaccessible region where no liquid
  state point can be found.
}
\label{ordermap}
\end{figure}

Fig.~\ref{ordermap} shows the isotherms from Fig.~\ref{qt-d} in the
ED order map representation. For both the 1SRP and 2SRP, the state points
fall on a two-dimensional region, i.e., $t$ and $Q_{6s}$ can be changed
independently. As is the case of silica and water
\cite{jeffrey01,shell02}, we also find an inaccessible region in the
ED order map where no liquid state points can be found. The structural
anomalies correspond to the section of the isotherms in Fig.~\ref{qt-d}
where both $Q_{6s}$ and $t$ decrease upon compression.  Only for the
2SRP we find such anomalies, as has been observed in both silica and
water. The state points corresponding to the structural anomalies do not
fall, strictly speaking, on a single line delimiting the inaccessible
region in the ED order map (Fig.~\ref{ordermap}(b)).  Instead, these state
points form a narrow stripe region, resembling water ED order map much more
closely than ED order map found for silica~\cite{shell02}.

\begin{figure}
\includegraphics[width=10cm]{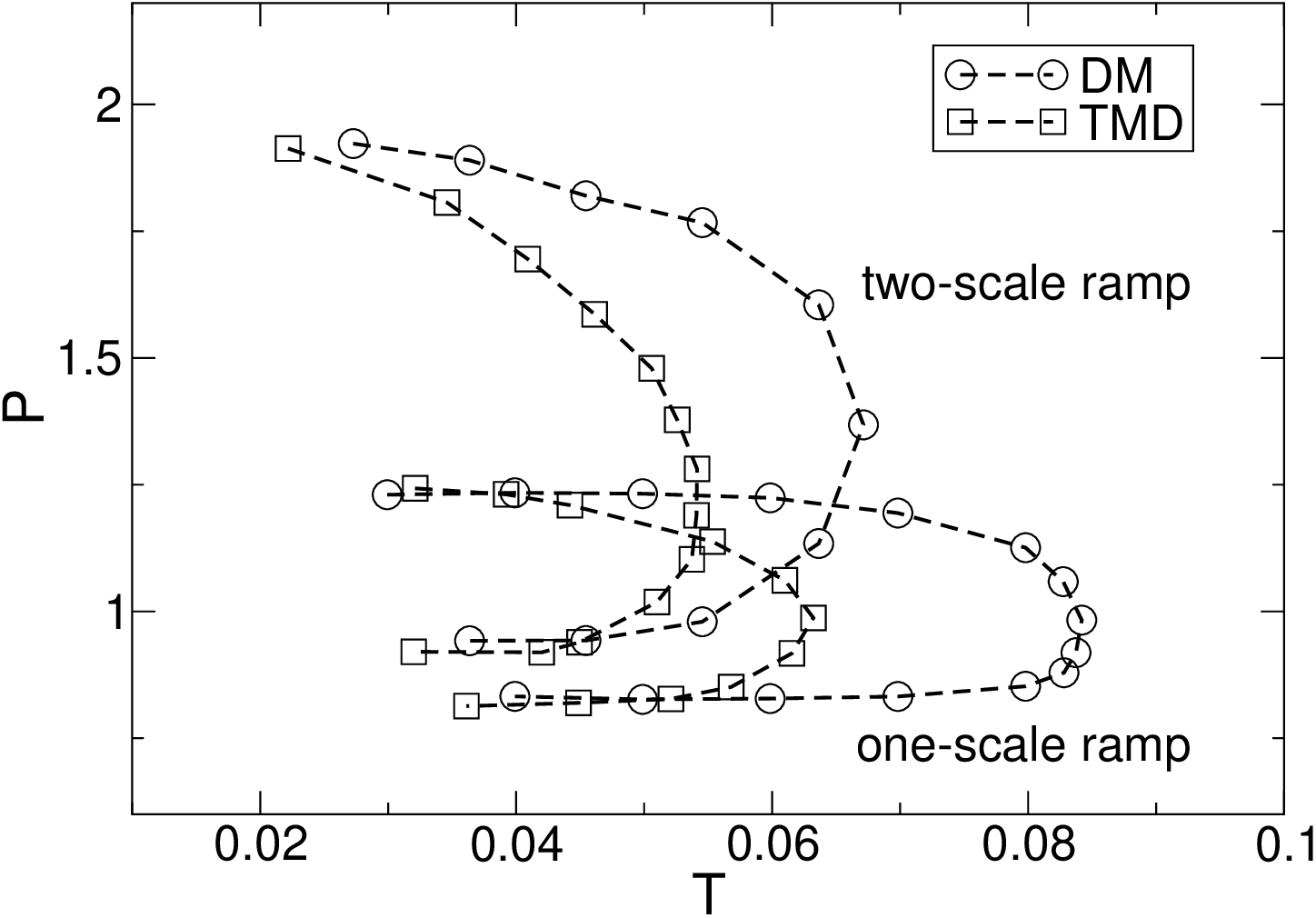}
\caption{ Temperature of maximum density (TMD) and diffusivity minima
     and maxima (DM) lines for the two potentials in the $P-T$ plane. The region of
     diffusion anomaly is defined by the loci of DM inside which the diffusivity increases with
     density. The thermodynamically anomalous region is defined by 
     TMD line, inside which the density
     increases when the system is heated at constant pressure. 
}
\label{phase}
\end{figure}

\begin{figure}
\includegraphics[width=12cm]{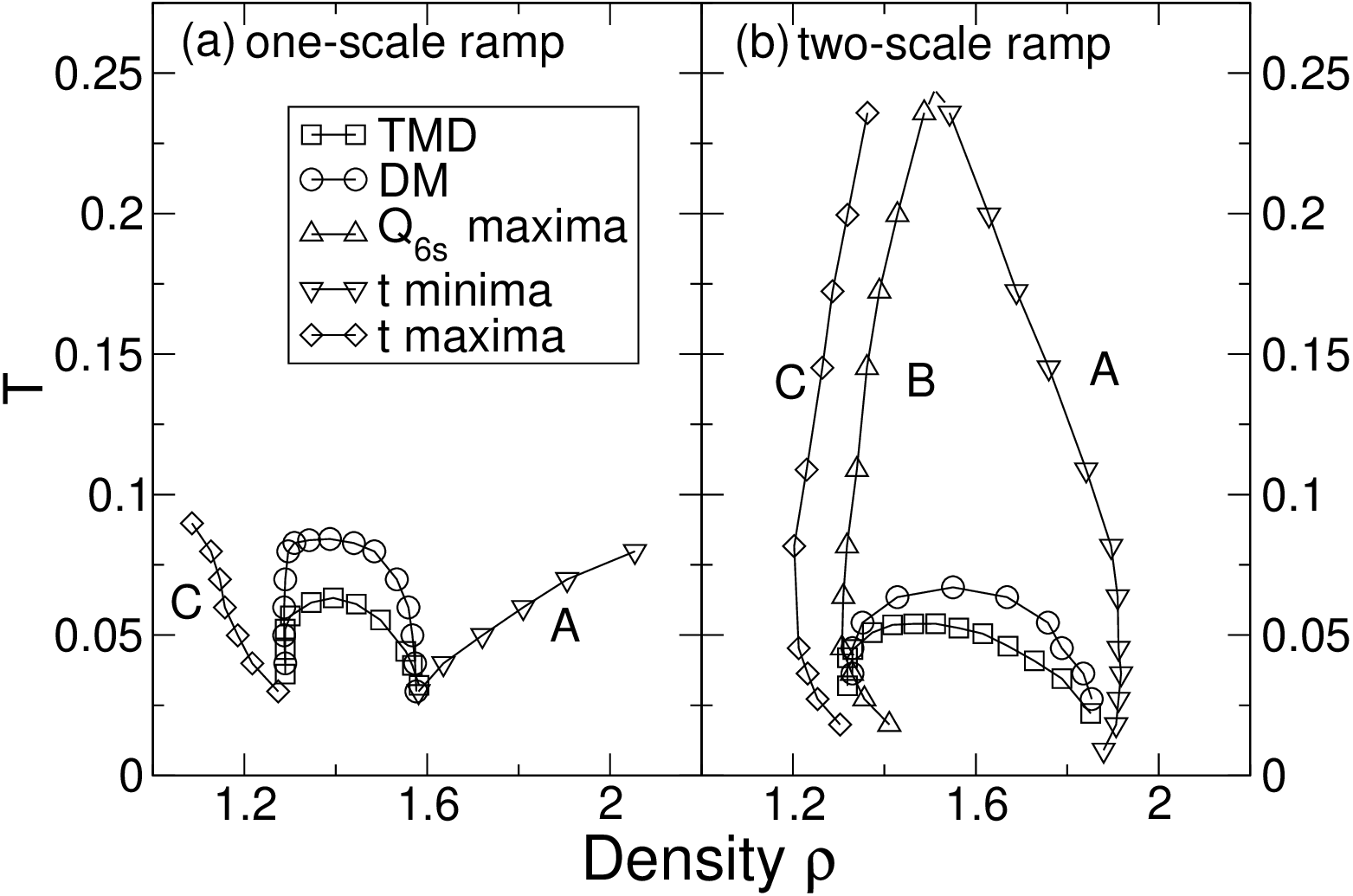}
\caption{
Relationship between the structural order and the density and
 diffusion anomalies. (a) For the one-scale ramp liquid, the open
 region bounded by the loci of $t$ minima (curve A) and $t$ maxima (curve C)
[see Fig.~\ref{ordermap}(a)] defines the structurally anomalous region. This
region contains the diffusion anomaly (delimited by the DM) and the
density anomaly (delimited by the TMD). (b)
For the two-scale ramp liquid, the structurally anomalous region
can be defined by the region between the loci of $t$ minima (curve A)
 and by the loci of either $t$ maxima (curve C) or $Q_{6s}$ maxima
(curve B) [see Fig.~\ref{ordermap}(b)]. The structurally anomalous 
region contains the diffusion anomaly
region which also contains the density anomaly region.
Only the two-scale ramp potential shows the same relation between the
structural, diffusion, and density anomaly regions as observed in water.}
\label{qttmd}
\end{figure}

Next, we discuss the regions in the phase diagram where the structural,
dynamical, and thermodynamic anomalies occur. In water,
the region in the $T-\rho$ or $P-T$ plane corresponding to
both structural anomalies contains the region corresponding to the
diffusion anomaly which in its turn contains the region corresponding to
the density anomaly. In silica, the diffusion anomaly region contains
the structural anomaly region, which also contains the density anomaly
region.
Fig.~\ref{phase} shows the temperature of
maximum density (TMD) line and the diffusivity maxima/minima (DM) line. The TMD
defines the boundary where the density anomaly occurs while the DM
defines the boundary where the diffusion anomaly occurs.  
By definition, the
structural anomaly region in the $T-\rho$ plane is delimited by the
location in the $T-\rho$ plane of the $Q_{6s}$ maximum and the $t$
minimum.  
Fig.~\ref{qttmd}(b) shows that the relation among the
regions of various anomalies for the 2SRP is the same as in the case of water
\cite{jeffrey01}, i.e., the structural anomaly region contains the
diffusion anomaly region which also contains the density anomaly region.

For the 1SRP, there is no clear $Q_{6s}$ maximum (i.e., $Q_{6s}$ shows
no anomaly), so we are not able to identify a structural
anomaly region.  However, in this case the structural anomaly can be
identified by those state points along an isotherm where $t$ decreases
upon compression. Therefore, the maximum and minimum values of $t$ in
Fig.~\ref{ordermap}(a) define the structural anomaly
region. Fig.~\ref{qttmd}(a) shows that with this definition the
structural anomaly region contains the diffusion anomaly region which
also contains the density anomaly region.
We show in Fig.~\ref{qttmd}(b) that for the 2SRP, defining the
structural anomaly boundaries using the extrema of $t$ does not alter
the relationship between the regions of various anomalies. Furthermore, comparing
Figs.~\ref{qttmd}(a) and \ref{qttmd}(b), we observe that the effect of
reducing the hard core distance $\sigma_0$ is to open the structural
anomaly region (curves C and A).

In summary, we find that the 2SRP shows not only density and diffusion
anomalies but also the same structural anomalies found in
tetrahedral liquids such as silica and water. Furthermore, the 2SRP
also shows the same relation among structural, dynamical and density
anomalies. Our finding suggests that the water-like relationship between
structural order and anomalies may be due to the presence of
two different repulsive length scales.  When eliminating the hard core
interaction, we find no water-like relation between structure and
dynamics. This suggests that the ratio between the two length scales in
the 2SRP, $\sigma_1/\sigma_0$, is the relevant variable in the interaction potential.

This work is an outgrowth of fruitful interactions with S. Chatterjee 
and P. G. Debenedetti. We thank P. G. Debenedetti for reading the manuscript 
and many critical suggestions. We also thank NSF grants
CHE~0096892 and CHE~0404699 for support and Yeshiva University
for location of CPU time.

\newpage

\newpage

\newpage

\newpage

\newpage

\begin{thebibliography}{LL}


\bibitem{angellPCCP}
C.A. Angell {\it et al.}, 
Phys. Chem. Chem. Phys. {\bf 2}, 7549 (2000).

\bibitem{phos}
Y. Katayama {\it et al.}, Nature  {\bf 403}, 170 (2000).
 
\bibitem{saika}
I. Saika-Voivod {\it et al.}, Phys. Rev. E  {\bf 63}, 011202 (2001).

\bibitem{sri}
S. Sastry and  C. A. Angell, Nature Materials {\bf 2}, 739 (2003).

\bibitem{poole}
P. H. Poole {\it et al.},
Nature {\bf 360}, 324 (1992).

\bibitem{pablo91}
P. G. Debenedetti {\it et al.}, J. Phys. Chem. {\bf 95}, 4540 (1991).

\bibitem{stillinger97} F. H. Stillinger and D. K. Stillinger, Physica A
  {\bf 244}, 358 (1997).

\bibitem{Jagla99} E. A. Jagla, J. Chem. Phys. {\bf 111}, 8980 (1999);   
 Phys. Rev. E {\bf 63}, 061509 (2001).
 
\bibitem{Sadr98} M. R. Sadr-Lahijany {\it et al.},
Phys. Rev. Lett. {\bf 81}, 4895 (1998); Phys. Rev. E {\bf 60},
6714 (1999).

\bibitem{Scala01} A. Scala et al.,
J. Stat. Phys. {\bf 100}, 97 (2000); Phys. Rev. E {\bf 63}, 041202
(2001).

 
\bibitem{Stell72} P. C. Hemmer and G. Stell, Phys. Rev. Lett.  {\bf 24}, 1284 (1970).


\bibitem{franzese} G. Franzese {\it et al.},
Nature {\bf 409}, 692 (2001); Phys. Rev. E {\bf 66}, 051206 (2002).

\bibitem{predpreprint}
P. Kumar {\it et al.}, Phys. Rev. E (in press)
cond-mat/0411274.

\bibitem{jeffrey01}
J. R.\ Errington and P. G.\ Debenedetti,
Nature {\bf 409}, 318 (2001).

\bibitem{errington03}
J. R. Errington {\it et al.},
J. Chem. Phys. {\bf 118}, 2256 (2003).

\bibitem{chau98}
P. L. Chau and A. J. Hardwick,
Mol. Phys. {\bf 93}, 511 (1998).

\bibitem{torquato00}
S. Torquato {\it et al.},
Phys. Rev. Lett. {\bf 84}, 2064 (2000).

\bibitem{shell02}
M. S. Shell {\it et al.},
Phys. Rev. E. {\bf 66}, 011202 (2002).

\bibitem{truskett00}
T. M. Truskett {\it et al.},
Phys. Rev. E. {\bf 62}, 993 (2000).

\bibitem{steinhardt83}
P. J. Steinhardt {\it et al.},
Phys. Rev. B. {\bf 28}, 784 (1983).

\bibitem{huerta04}
A. Huerta {\it et al.}, J. Chem. Phys. {\bf 120}, 1506 (2004).

\bibitem{footnoteDiffinQ6} 
In Refs.~\cite{errington03,torquato00,truskett00}, $\overline{Y}_{\ell
m}(\theta,\varphi)$ denotes the average over {\it all} the bonds in the
{\it system}, and the orientational order is defined for the whole system, and
not just for single molecules.  However, for water and silica
\cite{jeffrey01,shell02} the orientational order is defined for each 
molecule and its average is used.


\end{thebibliography}
\end{document}